\documentstyle[12pt,epsfig]{article}
\begin{document}
\newcommand{\newc}{\newcommand}
\newc{\R}{$R$}
\newc{\charginom}{M_{\tilde \chi}^{+}}
\newc{\mue}{\mu_{\tilde{e}_{iL}}}
\newc{\mud}{\mu_{\tilde{d}_{jL}}}
\newc{\beq}{\begin{equation}}
\newc{\eeq}{\end{equation}}
\newc{\barr}{\begin{eqnarray}}
\newc{\earr}{\end{eqnarray}}
\newc{\ra}{\rightarrow}
\newc{\lam}{\lambda}
\newc{\eps}{\epsilon}
\newc{\gev}{\,GeV}
\newc{\tev}{\,TeV}
\newc{\eq}[1]{(\ref{eq:#1})}
\newc{\eqs}[2]{(\ref{eq:#1},\ref{eq:#2})}
\newc{\etal}{{\it et al.}\ }
\newc{\Hbar}{{\bar H}}
\newc{\Ubar}{{\bar U}}
\newc{\Dbar}{{\bar D}}
\newc{\Ebar}{{\bar E}}
\newc{\eg}{{\it e.g.}\ }
\newc{\ie}{{\it i.e.}\ }
\newc{\cf}{{\it c.f.}}
\newc{\nonum}{\nonumber}
\newc{\sect}[1]{\ref{sec:#1}}
\newc{\labsec}[1]{\label{sec:#1}}
\newc{\lab}[1]{\label{eq:#1}}
\newc{\lle}[3]{L_{#1}L_{#2}\Ebar_{#3}}
\newc{\lqd}[3]{L_{#1}Q_{#2}\Dbar_{#3}}
\newc{\udd}[3]{\Ubar_{#1}\Dbar_{#2}\Dbar_{#3}}
\newc{\dpr}[2]{({#1}\cdot{#2})}
\newc{\rpv}{{\not \!\! R_p}}
\newc{\emis}{{\not \!\! E}}
\newc{\rpvm}{{\not \! R_p}}
\newc{\rp}{$R_p$}
\newc{\gsim}{\stackrel{>}{\sim}}
\newc{\lsim}{\stackrel{<}{\sim}}
\newc{\pbinv}{\,pb^{-1}}
\newc{\shat}{{\hat s}}

\newcommand{\chargino}{{\tilde \chi}^{+}}
\newcommand{\charginol}{{\chi}^{+}_{l}}
\newcommand{\neutralino}{{\tilde \chi}^{0}}

\newcommand{\ffig}[4]{\begin{figure}[htbp]\vfill\begin{center}
\mbox{\epsfig{figure=#1,height=#2}}\caption{#3}\label{#4}
\end{center}\vfill\end{figure}}

\newcommand{\wfig}[4]{\begin{figure}[htbp]\vfill\begin{center}
\mbox{\epsfig{figure=#1,width=#2}}\caption{#3}\label{#4}
\end{center}\vfill\end{figure}}

\newcommand{\pwfig}[4]{\begin{figure}[h]\vfill\begin{center}
\mbox{\epsfig{figure=#1,width=#2}}\caption{#3}
\end{center}\vfill\end{figure}}

\newcommand{\lwfig}[4]{\begin{figure}[h]\caption{#3}\label{#4}\end{figure}}

\setcounter{figure}{0}

\title{High $Q^2$-Anomaly at HERA and Supersymmetry} 
\author{H. Dreiner$^1$ and P. Morawitz$^2$}
\date{{\small $^1$ Rutherford Lab., Chilton, Didcot, OX11 0QX, UK\\
$^2$ Imperial College, HEP Group, London SW7 2BZ, UK}}
\maketitle


\begin{abstract}
We discuss the recently observed excess of high lepton momentum transfer, $Q^2$,
neutral current deep inelastic scattering events at HERA in the light of
supersymmetry with broken R-parity. We find more than one possible solution. We
consider the possibilities for testing these hypotheses at HERA, the Tevatron
and at LEP. One lepton-number violating operator can account for both the HERA
data and the four-jet anomaly seen by ALEPH at LEP. 
\end{abstract}

\newpage

\section{Introduction}
Recently both experiments at HERA have quoted an excess at high lepton momentum
transfer, $Q^2$, in their neutral current deep inelastic scattering (NC DIS)
data \cite{H1,ZEUS}. For $Q^2> 15,000 \gev^2$, $H1$ have found 12 events where
they expect $\approx5$ from the Standard Model (SM) \cite{H1}. For
$Q^2>20,000\gev^2$ $ZEUS$ observe 5 events where they expect $\approx2$ from
the SM \cite{ZEUS}. In addition, $H1$ have seen a slight excess in their
charged current deep inelastic scattering (CC DIS) \cite{H1}. Due to the low
rate, we find it is still too early to comment on this. The NC DIS 
discrepancies could be the first hint of physics beyond the SM \cite{prev}.
There have been three main suggestions predicting an excess of events at high
$Q^2$ at HERA: contact interactions \cite{eichten,rucklcontact,haberl}, 
leptoquarks \cite{ruckl,lqws}, and supersymmetry with broken R-parity
\cite{hewett,jonherbiws,jonherbi,kon1}.

If quarks (or leptons) have a substructure this could lead to an excess of
events at high values of $Q^2$ at colliders. A four-fermion operator based
analysis for testing effective theories at $e^+e^-$-colliders was first
proposed in Ref.\cite{eichten}. This analysis of contact interactions was
generalised to $eP$ deep inelastic scattering in \cite{rucklcontact} where the
operators are of the form $eeqq$. Contact interactions are a parametrisation of
unknown physics at a higher energy scale ($\Lambda\gg\sqrt{s}$)
via gauge-invariant but non-renormalisable operators. At HERA, they can lead to
an increase or a depletion of high $Q^2$ events, depending on the sign of the
interference with the SM-interaction, which is a free parameter. These events
should not show any peak-like structure, since the HERA energy is well below
any supposed resonances. At HERA, one can attempt to extract the unknown scale
of the new physics. This analysis is typically performed one-operator at a
time. But it is also possible to consider combined effects \cite{haberl}. This
should be done with the present HERA data, but we do not further consider this
possibility here.

Both leptoquarks and supersymmetry with broken R-parity predict s-channel
resonances in positron-proton scattering. In Grand Unified Models, the 
leptoquarks are typically heavy and well beyond the reach of colliders 
\cite{gut}. However, there is a set of models which predict 
low-energy leptoquarks \cite{lightlq}. A systematic search for such leptoquarks
was first proposed in Ref.\cite{ruckl} and subsequently much effort has been
dedicated at HERA to this search \cite{lqws}. Within supersymmetry, we
naturally expect all new states to have masses between $100\gev$ and $1\tev$ in
order to maintain the solution to the gauge hierarchy problem
\cite{gaugehierarchy}. Depending on the magnitude of the Yukawa coupling and
the remaining supersymmetry spectrum, squarks can behave just like leptoquarks
at HERA. But squarks can naturally have further decay modes and their
discussion is thus more general. In the following, we focus on the implications
of the recent HERA data for supersymmetry with broken R-parity. When relevant
we shall emphasise below where the leptoquark description is distinct.

In Section \sect{rpv} we briefly review supersymmetry with broken R-parity and
focus on the relevant terms for HERA. In Section \sect{bounds} we review the 
indirect bounds on these operators. In Section \sect{signal} we determine
which R-parity violating operators could possibly lead to the observed excess 
at HERA while still being consistent with the existing indirect bounds. In 
Sections \sect{hera} through \sect{fermilab} we discuss how these operators 
could be tested in future runs at HERA, as well as with present and upcoming 
data at LEP and at the Tevatron. In Section \sect{concl} we present our 
conclusions. In the Appendix we have collected some of the relevant formula 
for completeness.

\section{Supersymmetry with Broken R-Parity at HERA}
\label{sec:rpv}
When minimally extending the particle content of the SM to incorporate 
supersymmetry one must add an extra Higgs $SU(2)_L$ doublet and then double
the particle content. The most general interactions of these particles
consistent with supersymmetry and $SU(3)_c\times SU(2)_L\times U(1)_Y$ gauge
symmetry are those of the minimal supersymmetric standard model (MSSM)
\cite{MSSM} as well as the superpotential terms\footnote{There is the further
term $\kappa_iL_iH_u$ which violates lepton number. If the soft-breaking terms
are universal it can be rotated away \cite{hallsuzuki}.}\cite{rpsuper}
\beq
\lam_{ijk}\lle{i}{j}{k}+\lam'_{ijk}\lqd{i}{j}{k}+\lam''_{ijk}\udd{i}{j}{k}.
\lab{rpv}
\eeq
Here $L$ ($Q$) is the lepton (quark) doublet superfield, and
$\Dbar,\Ubar$ ($\Ebar$) are the 
down-like and up-like quark (lepton) singlet superfields, respectively. 
$i,j,k=1,2,3$ are generation indices.
The last two sets of terms in \eq{rpv} lead to rapid proton decay 
\cite{vissani}. The solution of this problem in the MSSM is to impose 
R-parity, 
\beq
R_p=(-{\bf 1})^{3B+L+2S},
\eeq
a multiplicative discrete symmetry. Here $B$ denotes baryon number, $L$ lepton
number\footnote{This should not lead to any confusion with the lepton doublet 
superfield, $L$, of Eq.\eq{rpv}.} and $S$ the spin of a field. All SM fields 
have $R_p=+1$; their
supersymmetric partners have $R_p=-1$. This solution is not unique. There are 
many models which protect the proton but allow a subset of the terms in 
\eq{rpv} \cite{hallsuzuki,aliherbi,models}. This subset can be as small as two
operators even for a gauge symmetry \cite{aliherbi}. All these alternative
solutions are denoted ``R-parity violation''. In the following, we shall focus
on the subset of the operators \eq{rpv}
\beq
\lam'_{1jk}\lqd{1}{j}{k}.
\lab{rpvlqd}
\eeq
At HERA these operators can lead to resonant squark production 
\barr
e^+ + {\bar u}_j &\ra& {\tilde {\bar d}}_k, \lab{squarkprodd}\\
e^+ + {d}_k &\ra& {\tilde {u}}_j .
\lab{squarkprodu}
\earr
This was first proposed by J. Hewett in Ref.\cite{hewett} where she considered
the direct R-parity violating decay of the squark to the initial state. The
processes \eqs{squarkprodd}{squarkprodu} were discussed in more detail in 
\cite{jonherbiws,jonherbi} where the squark cascade decays via a photino were
included. This enables the distinction from a leptoquark, depending on the
supersymmetric spectrum. The full mixing in the neutralino sector was first
considered in \cite{peterherbi} for R-parity violation at HERA. For a scalar 
top quark
($j=3$), the decays via the neutralino are kinematically forbidden. This 
scenario was discussed in considerable detail by T. Kon \etal \cite{kon1,kon}. 
The
squark decays for the full gaugino mixing including the chargino were discussed
by E. Perez \cite{emmanuelle} and in a shorter version in \cite{meheraws}. 
This enabled a full classification of the final state topologies at HERA.

Making use of all this work, we wish to determine the size and generation 
structure $j,k$ of the coupling $\lam'_{1jk}$ of \eq{rpvlqd} which is 
preferred by the HERA data. We would also like to consider the best estimate 
for the squark mass. However, in order to do that we must first consider the 
indirect bounds on the couplings $\lam'_{1jk}$.

\section{Previous Indirect Bounds}
\labsec{bounds}
The operators \eq{rpvlqd} can contribute to several processes with initial and 
final state SM particles via the exchange of virtual sleptons or squarks. Since
to date all such processes have been observed in agreement with the SM, this 
leads to upper bounds on all the couplings which we summarise in 
Table~\ref{tab:rpv.bounds}.
\begin{table}[t]
\centerline{\begin{tabular}{|lllllllll|}
\hline
$\lam'_{111}$&$\lam'_{112}$&$\lam'_{113}$&$\lam'_{121}$&$\lam'_{122}$&$
\lam'_{123}$&$
\lam'_{131}$&$\lam'_{132}$&$\lam'_{133}$\\\hline
$0.001^{(a)}$&$0.03^{(b)}$&$ 0.03^{(b)} $&$  0.06^{(c)}$&$0.06^{(d)}$ 
&$0.26^{(e)}$&$0.06^{(c)}$&$0.63^{(f)}$&$0.002^{(d)}$
\\ 
$0.004$&$0.06$&$ 0.06 $&$  0.13$&$0.087$ 
&$0.55$&$0.13$&$1.3$&$0.003$ \\ \hline
\end{tabular}}
\caption[yy]{\label{tab:rpv.bounds}
{\small Indirect bounds on first lepton generation operators $LQ\Dbar$. The
first line is the bound for a scalar fermion mass of $100\gev$ the second line
for $210\gev$. The bounds derive from the following physical processes: $^{(a
)}$ neutrinoless double beta decay \cite{klapdor}, $^{(b)}$ charged current
universality \cite{barger,davidson,pdg}, $^{(c)}$ atomic parity violation
\cite{davidson,langacker,pdg}, $^{(d)}$ $\nu_e$ mass \cite{tata}, $^{(e)}$
forward backward asymmetry \cite{barger}, and $^{(f)}$ D-decays \cite{D1}. The
bounds from $^{(b)}$, $^{(c)}$, $^{(e)}$, and $^{(f)}$ scale linearly with the
squark mass $({\tilde M}_{\tilde q}/100)\gev$. The bound from $^{(d)}$ scales
with the square root $\sqrt{{\tilde M}_{\tilde q} /100\gev}$. The bound
$^{(a)}$ scales as $({\tilde M}_q/100\gev)^2\sqrt{ {\tilde M}_{\tilde
g}/100\gev}$. We have conservatively estimated the gluino mass at ${\tilde
M}_{\tilde g}=1\tev$. We have given the 1 sigma bounds.}} 
\end{table}
We give the bounds for a scalar fermion mass of $100\gev$, which is the 
standard, and $210\gev$, which is what we require below. These two bounds are
related via scaling properties which are explicitly given in the table caption.
We have updated some previous bounds using more recent data from the PDG 
\cite{pdg}. The bound from atomic parity violation is obtained using the ``weak
charge'', $Q_W$, \cite{pdg} and was not previously applied to R-parity
violation \cite{davidson}. All but two of the bounds are well below the 
electromagnetic coupling.

There is a further set of stringent bounds from the decay $K\ra\pi\nu\nu$ 
\cite{kpinunu}.
\barr
\lam'_{1jk}&<0.012,\quad j=1,2,\quad k=1,2,3;\quad &{\tilde M}=100\gev,\\
\lam'_{132}& \hspace{-4.85cm}
<0.19, &{\tilde M}=100\gev.
\earr
These limits scale linearly with the squark mass. However, these bounds are
model dependent. If the absolute mixing of the quarks (not squarks) is purely
in the down-quark sector then these bounds apply. If it is purely in the
up-quark sector then these bounds revert to those of Table 1. In order to keep
an open mind to all {\it possible} solutions we do not further consider these
model dependent bounds. 

If there is a charged, first or second generation doublet slepton (${\tilde 
e}_L,\,{\tilde \mu}_L$) lighter than the top quark there is an additional bound
from top quark decays \cite{kpinunu} 
\beq
\lam'_{132}< 0.4, \quad M_{\tilde \ell}=100\gev,
\eeq
which is more strict than the one given in Table 1.

$H1$ have performed a direct search for supersymmetry with broken R-parity
\cite{H1.photino.search}. The bounds from this search depend on the neutralino
mass. For a neutralino mass $M_{{\tilde\chi}^0_1}=40\gev<M_{{\tilde q}}$ they 
obtain
\barr
\lam'_{1j1}&<0.2,\quad M_{\tilde q}&=200\gev,\quad M_{{\tilde\chi}^0_1}=40,\\
\lam'_{132}&<0.22,\quad M_{\tilde q}&=150\gev,\quad M_{{\tilde\chi}^0_1}=80.
\earr
In particular the last bound is significantly better than the indirect bound
of Table 1, but is also model dependent.

\section{HERA's high $Q^2$ excess interpreted as s-channel squark production}
\label{sec:signal}
\subsection{R-parity Violating Squark Decays}
\begin{figure}
\begin{center}
{\psfig{figure=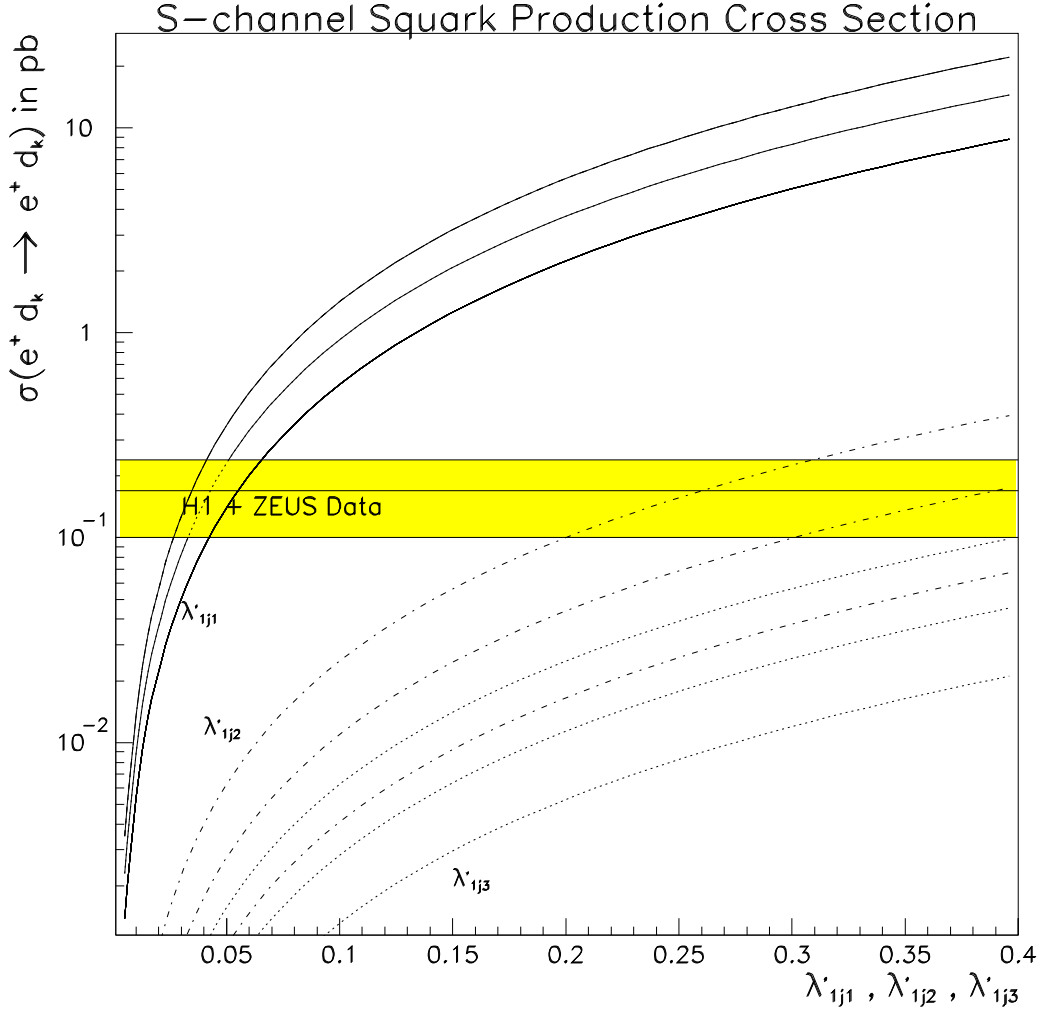,height=15cm}}
\end{center}
\caption[xx]{\label{sup.xsec}{\small Excess cross-section $\sigma(e^+ d_k \ra e^+ d_k)$
for a s-channel $\tilde u_j$-squark resonance  for $Q^2>20,000\gev$ as a function of the coupling $\lambda'_{1jk}$ for 
the three values of $M_{\tilde u_j}=200,210,220\gev$ (top to bottom). The 
solid lines show the cross-sections for a non-zero coupling $\lambda'_{1j1}$ 
(\ie the d valence quark contribution), while the dashed-dotted and dotted 
lines show the cross-sections for the non-zero couplings $\lambda'_{1j2},$ 
and $\lambda'_{1j3}$ (\ie the $s$, and $b$ sea quark contributions), 
respectively. Here, we assume that the only allowed squark decay mode is ${
\tilde {u}}_j \ra e^+ +{d}_k$. The hatched region shows the high $Q^2$ excess 
cross-section $\sigma_{ex}=(0.17 \pm 0.07)\,pb$.}}
\end{figure}

\begin{figure}
\begin{center}
{\psfig{figure=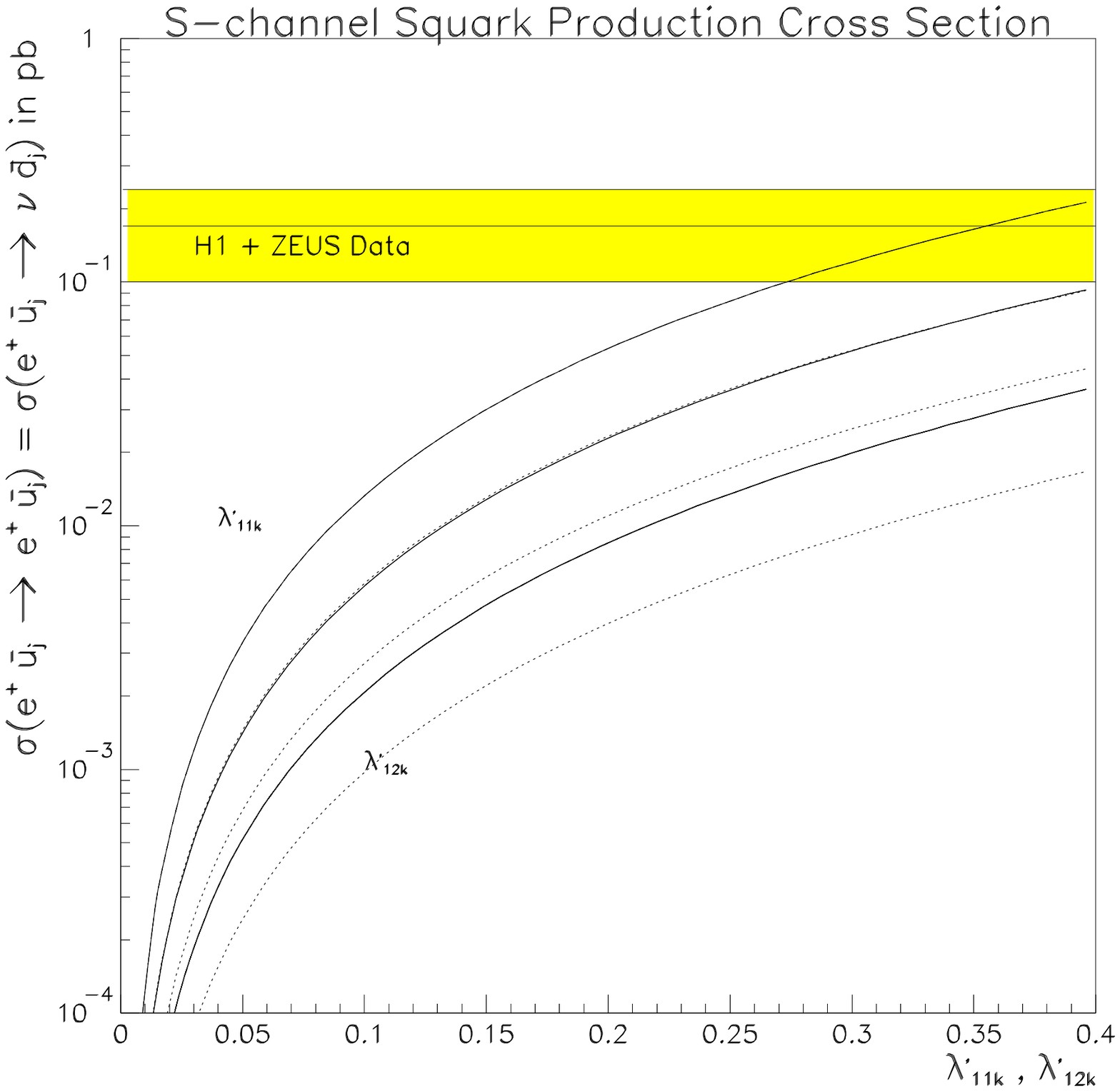,height=15cm}}
\end{center}
\caption[xx]{\label{sdown.xsec} {\small Excess cross-section $\sigma(e^+ {\bar u}_j \ra e^+
{\bar u}_j) = \sigma(e^+ {\bar u}_j \ra \nu {\bar d}_j)$ for a s-channel 
$\tilde {\bar d}_k$-squark resonance  for
$Q^2>20,000\gev$ as a function of the coupling $\lambda'_{1jk}$ for the three 
values of $M_{\tilde d_k}=200,210,220\gev$ (top to bottom). The solid lines 
show the cross-sections for a non-zero coupling $\lambda'_{11k}$ (\ie the 
u-bar sea quark contribution), while the dotted lines show the cross-sections
for the non-zero coupling $\lambda'_{12k}$ (\ie the c-bar sea quark
contributions), respectively. Here, we assume that the only allowed 
squark decay modes are ${\tilde {\bar d}}_k\ra e^+ + {\bar u}_j$ and 
${\tilde{\bar
d}}_k\ra{\bar\nu}+{\bar d}_j$. The hatched region shows the high
$Q^2$ excess cross-section $\sigma_{ex}=(0.17 \pm 0.07)\,pb$.}}
\end{figure}

We would now like to interpret the observed excess at HERA in terms of
supersymmetry with R-parity violation. For this, we combine the $H1$ and $ZEUS$
data in order to compare it more easily with different R-parity violating 
models. For $Q^2>20,000\gev$, $H1$ and $ZEUS$ see a total of 10 events 
\cite{H1,ZEUS}, where $4.08\pm0.36$ events are expected from SM contributions
\cite{us}. The total integrated luminosity of the two experiments is
34.3$\pbinv$, which translates into an excess cross-section over the SM
expectation of 
\barr
\sigma_{ex}(Q^2>20,000\gev)&=&(0.17 \pm 0.07(stat))\, pb.
\lab{sig.excess}
\earr
In Figures~\ref{sup.xsec} and \ref{sdown.xsec} we show this combined excess 
cross section as a horizontal band. 

The combined reconstructed values of the hypothetical squark mass $M_{\tilde q
}$ (where $M_{\tilde q}$ is related to  Bjorken-x by $M_{\tilde q}^2=x s$ and 
$s$ is the centre-of-mass energy squared) show some spread between the two 
experiments ($M_{\tilde q}=200\gev$ for $H1$, $M_{\tilde q}=220\gev$ for $ZEUS
$). Combining the two experiments, we obtain\footnote{The errors quoted on the
mass of the excess events ($M_e$) in Table 6 \cite{H1} is not the actual
error on the reconstructed mass of any supposed resonance. This error is most
likely larger as can be seen from the broad width of the distributions in Fig.
2b of Ref. \cite{H1}. Analogous, remarks apply to the ZEUS analysis
\cite{ZEUS}. We therefore feel justified in combining the data in this way.}
\barr
M_{\tilde q}&=& (210 \pm 20)\gev,
\lab{mass}
\earr
as our best estimate.

In order to determine the contribution from R-parity violation we consider one
of the  operators \eq{rpvlqd} at a time while assuming the others are 
negligible. For a given non-vanishing operator $\lam'_{1jk} \lqd{1}{j}{k}$, the
produced ${\tilde u}_j,$ or ${\tilde {\bar d}}_k$ squark can have many decay 
modes, depending on the mass spectrum of the supersymmetric particles 
\cite{jonherbiws,jonherbi,emmanuelle,meheraws,decaymodes}. For all SUSY
spectra, the squark can decay via the operator itself resulting in the
interactions
\barr
e^+ + {\bar u}_j &\ra& {\tilde {\bar d}}_k\ra e^+ + {\bar u}_j,\lab{squark1}\\
e^+ +{\bar u}_j&\ra&{\tilde{\bar d}}_k\ra{\bar\nu}+{\bar d}_j, \lab{squark2}\\
e^+ + {d}_k &\ra& {\tilde {u}}_j \ra e^+ +{d}_k,
\lab{squark3}
\earr
at HERA. The first and third have equivalent initial and final states to
NC DIS and can thus contribute to the observed excess. Analogously, the second
can contribute to CC DIS. Let us for now assume these 
are the only decay modes of the produced ${\tilde {u}}_j$ {\it or} ${\tilde {
\bar d}}_k$ squark. We then compute the production cross section using 
\cite{hewett,jonherbi}
and the {\it MRSG} structure functions \cite{mrs}. For 
\eq{squark1} 
and \eq{squark2} we include the extra contribution to the width for the other 
decay mode. The correction to the narrow width approximation by using the full 
resonance width is $\lsim10\%$.

In Figure~\ref{sup.xsec}, we plot the production cross section $\sigma(e^+ +{d}
_k \ra{\tilde {u}}_j \ra e^+ +{d}_k)$ for $Q^2>20,000\gev$ as a function of the
R-parity violating couplings $\lam'_{1j1},\,\lam'_{1j2},\,\lam'_{1j3}$. For
each coupling, the three curves are for the squark masses $ M_{{\tilde
u}_j}=200,\,210,$ and $220\gev$ from top to bottom, respectively. The cross
section is largest for $\lam'_{1j1}$ due to the incoming valence quark. The
other two cross sections are suppressed because of the significantly smaller
strange and bottom sea-quark structure functions $s(x),\,b(x)$. 

In Figure~\ref{sdown.xsec}, we analogously plot the cross sections $\sigma(e^+ 
+ {\bar u}_j \ra {\tilde {\bar d}}_k\ra e^+ + {\bar u}_j)$ as a function of the
 couplings $\lam'_{11k}$ and $\lam'_{12k}$ for the three squark mass values $M_
{{\tilde d}_k}=200,\,210,$ and $220\gev$. Here, we have included the full 
R-parity violating width from the two ${\tilde d}_j$ decay modes. These cross 
sections are both suppressed because of the small up and charm sea-structure 
functions ${\bar u}(x)$ and ${\bar c}(x)$, respectively, as well as the
increased width. There is no contribution for $\lam'_{13k}$ because of the
unknown and suppressed top-quark structure function, ${\bar t}(x)$.

We now compare the plotted cross sections with the hatched band of 
Eq.\eq{sig.excess}, and the bounds in the second row of 
Table~\ref{tab:rpv.bounds}. We obtain the following set of solutions 
where R-parity violation can explain the excess in the HERA data
\beq
\begin{array}{cccccccc}
(1) & \lam'_{121}&\approx&0.04, &\quad {\tilde c},&\quad M_{\tilde c}&=&
(210\pm
20)\gev, \\
(2) & \lam'_{123}&\gsim& 0.4, &\quad {\tilde c},&\quad M_{\tilde c}&<&
210\gev, \\
(3) & \lam'_{123}&\gsim& 0.4, & \quad{\tilde b}, &\quad M_{\tilde b}&<&
210 \gev,  \\
(4) & \lam'_{131}&\approx&0.04, &\quad {\tilde t}, &\quad M_{\tilde t}
&=& (210\pm20)\gev,  \\
(5) & \lam'_{132}&\gsim&0.3, &\quad {\tilde t},&\quad M_{\tilde t}&=&
(210\pm20)\gev. 
\end{array}
\lab{solutions}
\eeq
In each row we first present the approximate value of the required Yukawa 
coupling, then denote the produced squark and  the mass range of the squark 
which is viable.  Thus we obtain many
solutions; however, the preferred solutions which are well within the
constraints are (\ref{eq:solutions}.1), (\ref{eq:solutions}.4) and
(\ref{eq:solutions}.5) with a produced scalar charm quark and a scalar top
quark (``stop''), respectively. Before discussing the other squark decay modes,
we note that within supersymmetric unification scenarios there is a possibility
for a very light stop, possibly even the LSP \cite{stop}. For a stop-LSP, the
process (\ref{eq:squark3}) would be the only decay.

There is one possible further solution, which is intriguing. Combining the 
solutions (\ref{eq:solutions}.2) and (\ref{eq:solutions}.3), it is possible 
that HERA has produced two different squarks: a scalar charm {\it and} a
scalar bottom. The required coupling is then reduced by a factor $\sqrt{2}$
to $\lam'_{123}\approx0.28$, which is well away from the indirect bound. 
\barr
(7)\; \lam'_{123}&\!\!\!\!=0.28,\quad &{\tilde c}\;\&\;{\tilde b},\quad 
M_{\tilde q}\approx210\gev. \lab{special1} 
\earr

We make one further point \cite{zoltan}. The leading order QCD correction to
the resonant squark production is presently not known. On-shell squark
production is effectively a $2\ra1$ process for which the QCD corrections are
typically positive and large \cite{spira}. If this is indeed the case for 
resonant squark
production at HERA this would give additional lee-way in the Yukawa-coupling
for the more marginal solutions. The required coupling would be reduced to 
$\lam'/\sqrt{K}$, where $K$ is the QCD $K$-factor.

\begin{figure}
\begin{center}
{\psfig{figure=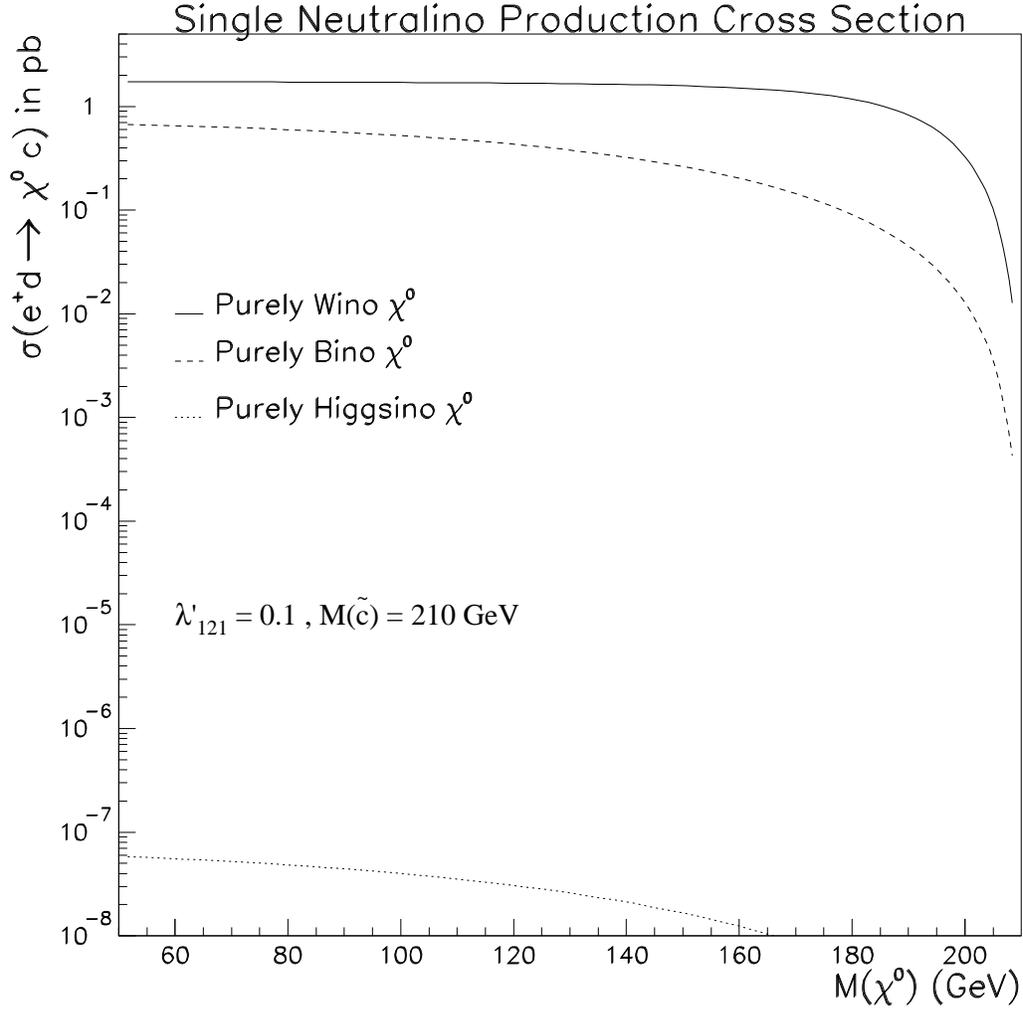,height=15cm}}
\end{center}
\caption[xx]{\label{neutralino.xsec} {\small Cross-section $\sigma(e^+ d \ra {\tilde \chi}^0 c)$ 
for single neutralino
production at HERA for the coupling $\lambda'_{121}=0.1$ and $M_{\tilde
c}=210\gev$ as a function of the neutralino mass. The solid, dashed and 
dotted lines show the cross-sections for a neutralino which couples purely 
Wino-, Bino-, or Higgsino-like, respectively.}}
\end{figure}

\begin{figure}
\begin{center}
{\psfig{figure=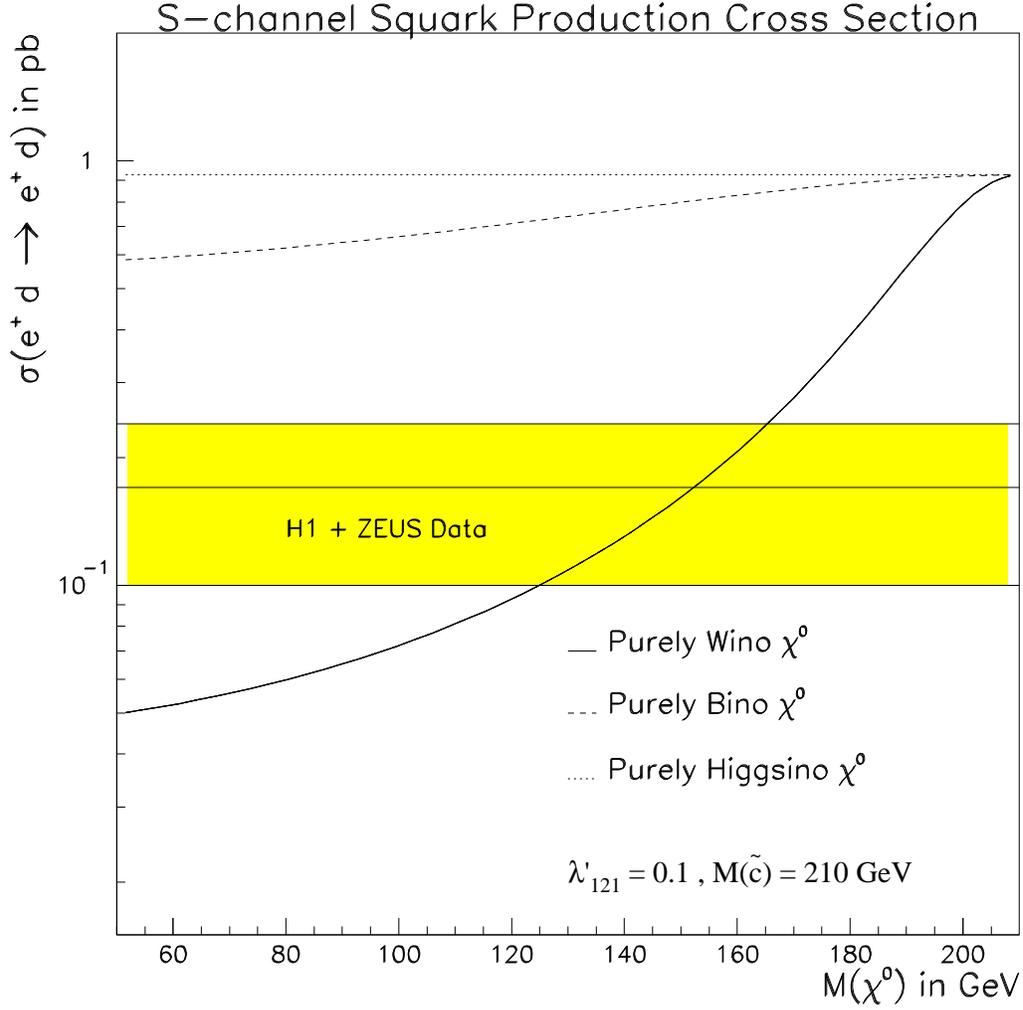,height=15cm}}
\end{center}
\caption[xx]{\label{sup.xsec.reduced} 
{\small Effect of the additional decay mode $\tilde c \ra {\tilde \chi}^0 c$ on the 
excess cross-section $\sigma_{ex}(e^+ d \ra e^+ d)$ for a s-channel $\tilde 
c$-squark resonance for $Q^2>20,000
\gev$, $\lambda'_{121}=0.1$ and $M_{\tilde c}=210\gev$, as a function of 
the neutralino mass. The hatched region shows the high $Q^2$ excess 
cross-section $\sigma_{ex}=(0.17 \pm 0.07)\,pb$.}}
\end{figure}

\begin{figure}
\begin{center}
{\psfig{figure=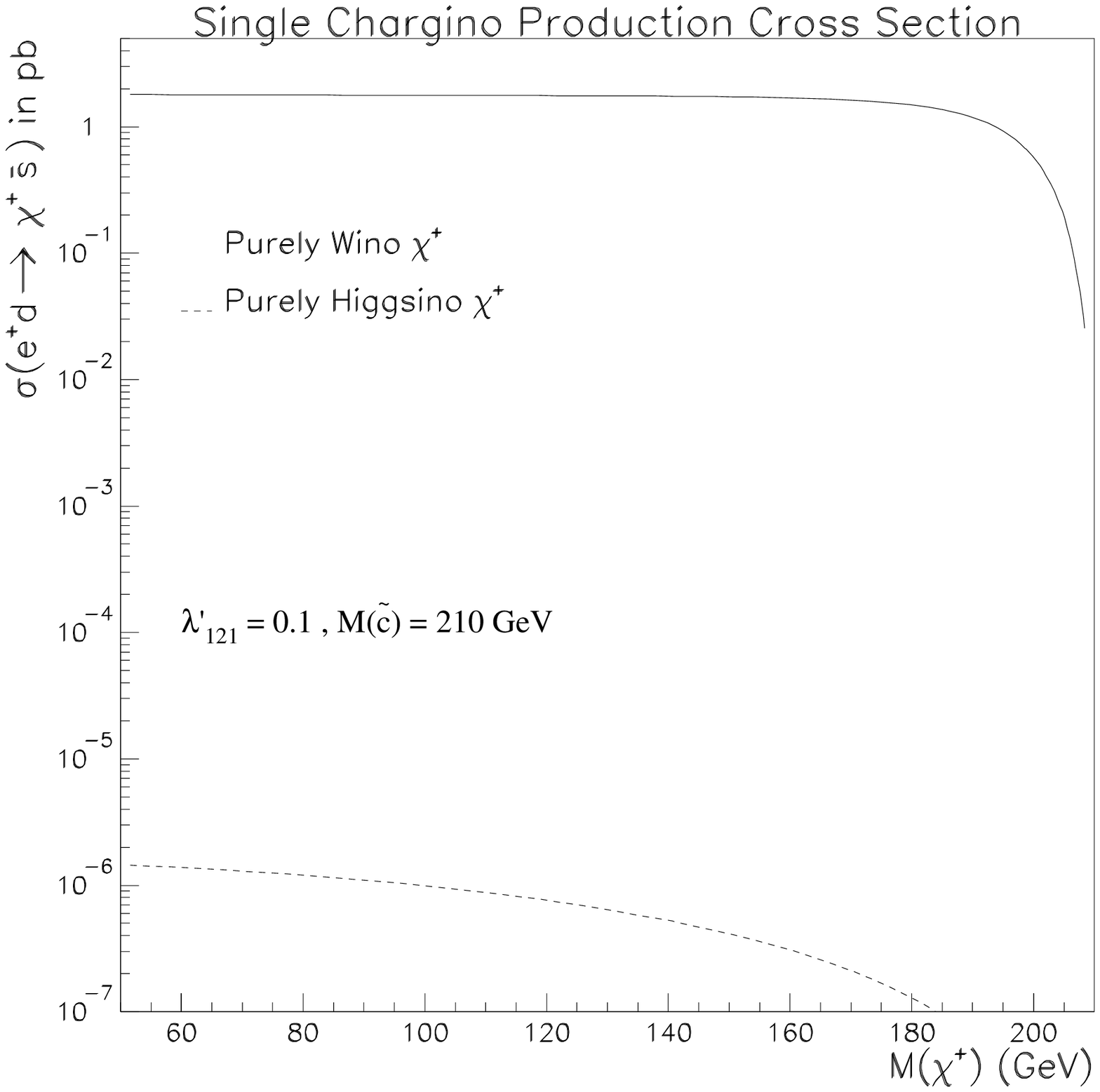,height=15cm}}
\end{center}
\caption[xx]{\label{chargino.xsec}{\small Cross-section $\sigma(e^+ d \ra {\tilde \chi}^+ s)$ 
for single chargino production at HERA for the coupling $\lambda'_{121}=0.1$ 
and $M_{{\tilde c}}=210\gev$ as a function of the chargino mass. The
solid and dashed lines show the cross-sections for a chargino which couples 
purely Wino-, or Higgsino-like, respectively.}}
\end{figure}

\subsection{Gaugino Decays of the Squark}
In this Subsection we wish to discuss the case where the produced squark is not
the LSP and can thus have further supersymmetric decay modes. (This section has
no analogy for leptoquarks.) As noted above, supersymmetry dramatically extends
the spectrum of the SM. It is way beyond the scope of this paper to perform a
systematic study of all the possible decay chains. Instead, we focus on
two specific cases which are well motivated by the renormalisation
group studies of the MSSM \cite{rgestudies}. (a) The lightest
neutralino, ${\tilde \chi}^0_1$, is lighter than the produced squark,
$M_{{\tilde \chi}^0_1}< M_{{\tilde u}_j},\,M_{{\tilde d}_k}$ and (b) The 
lightest chargino, ${\tilde \chi}^+_1$, is lighter than the produced squark,
$M_{{\tilde \chi}^+_1}< M_{{\tilde u}_j},\,M_{{\tilde d}_k}$. 

(a) If 
\beq
M_{{\tilde \chi}^0_1}< M_{{\tilde u}_j},\,M_{{\tilde d}_k},
\eeq
then we have the following additional interactions at HERA
\barr
e^++{\bar u}_j&\ra {\tilde {\bar d}}_k&\ra {\tilde\chi}_1^0 + {\bar d}_k, \\
e^++{d}_k&\ra {\tilde {u}}_j&\ra {\tilde\chi}_1^0 + {u}_j. 
\earr
In the second process, the case $j=3$ is suppressed by the large top-quark mass
and we do not further consider it. 

The production cross section depends on the admixture of the neutralino mass
eigenstate. In order to determine the cross section we focus on three special
limiting cases: a purely Wino-, a purely Bino- and a purely Higgsino-${\tilde
\chi}^0_1$. In Figure~\ref{neutralino.xsec} we plot the production cross
section $\sigma(e^++ {d}_k\ra {\tilde\chi}_1^0 +{u}_j )$ as a function of the
neutralino mass for these 3 special cases.\footnote{The single 
neutralino/chargino production cross-sections are quoted in the Appendix for 
completeness.}  We have fixed the squark mass at
$210\gev$ and the Yukawa coupling $\lam'_{1j1}=0.1$. For a purely
Higgsino-neutralino, the production cross section is very small ($<10^{-6}\,
pb$) and is not observable at HERA. For a gaugino-neutralino, the cross section
is $\sim1pb$ for neutralino masses as large as $150\gev$ or $180\gev$. With
the present luminosity, HERA could possibly have already produced several tens
of neutralinos, upto about 30 per experiment. The neutralino decay modes are
\cite{peterherbi} 
\beq
{\tilde \chi}^0_1\ra \left\{e^-u_j{\bar d}_k, \; e^+{\bar u}_j{d}_k, 
\; \nu_ed_j{\bar d}_k,\;{\bar \nu}_e{\bar d}_j{d}_k
\right\}.
\eeq
The branching fractions are not simply $0.25$ each for massless final states
but sensitively depend on the admixture of the neutralino \cite{peterherbi}. The 
most visible decay for a positron beam is ${\tilde \chi}^0_1\ra e^-u_j{\bar d}_
k$, \ie to the final state electron \cite{jonherbi}. This requires charge
identification of high $p_T$ electrons and can be searched for in the present
and upcoming data.

By including the neutralino decay of the squark we have increased the total 
decay width of the squark. This in turn reduces the resonant production cross 
sections plotted in Figure~\ref{sup.xsec}. In Figure~\ref{sup.xsec.reduced}, we
show this effect for the three neutralino admixtures and for $\lam'_{1j1}=0.1
$, $M({\tilde u}_j)=210\gev$. The change in the leptoquark-like cross section
(squark LSP) is negligible for the Higgsino-neutralino. For a Bino-neutralino,
the cross section can drop by about a factor two. For a Wino-neutralino the
production cross section can drop by more than an order of magnitude. 

If we include a Bino- or Wino-${\tilde\chi}_1^0$ which is lighter than the 
produced squark, we must reconsider the solutions presented in Section 4.1. For
the same coupling $\lam'_{1jk}$, the rate for the reactions
\eqs{squark1}{squark3} will go down when including the neutralino. This can be
compensated  by increasing the Yukawa coupling, if allowed. It is only for the
solutions (\ref{eq:solutions}.1,\ref{eq:solutions}.4,\ref{eq:solutions}.5) 
and (\ref{eq:special1}) that this is
possible. For example, for solution (\ref{eq:solutions}.1), producing a
scalar-charm quark, we can raise $\lam'_ {121}=0.04$ by a factor 2.5 to
$\lam'_{121}=0.1$. We then obtain a new solution for a purely Wino-${\tilde
\chi}_1^0$ with mass $M_{{\tilde\chi}_1^0}=125-165 \gev$. This can be seen in
Figure~\ref{sup.xsec.reduced}. The branching ratio for the direct decay 
${\tilde c}\ra e^+d$ is reduced to about 0.1. For a purely Bino-${ \tilde
\chi}_1^0$ only a modest increase in coupling can be allowed, otherwise the
suppression is not sufficient.

We can similarly consider the effects of the decay ${\tilde{\bar d}}_k\ra
{\tilde\chi}_1^0+{\bar d}_k$ on the production cross section $\sigma(e^+{\bar
u}_j\ra e^+ {\bar u}_j)$. The $SU(2)$ singelt $d$-squark only couples 
significantly to a purely Bino-${\tilde \chi}_1^0$. This only leads to marginal 
affects, as seen above. Except for the two squark solution \eq{special1}, these
solutions are already marginal and we do not further consider them.

(b) We can repeat the analogous analysis for a chargino with
\beq
M_{{\tilde \chi}^+_1}< M_{{\tilde u}_j},\,M_{{\tilde d}_k},
\eeq
in which case we obtain the additional interactions
\barr
e^++{\bar u}_j&\ra {\tilde {\bar d}}_k&\ra {\tilde\chi}_1^+ + {\bar u}_k, \\
e^++{d}_k&\ra {\tilde {u}}_j&\ra {\tilde\chi}_1^+ + {d}_j. 
\earr
Here we also allow for the case $j=3$. We consider the two limiting cases, where the
chargino is pure Wino-${\tilde \chi}_1^+$ or pure Bino-${\tilde\chi}_1^+$. In
Figure~\ref{chargino.xsec} we plot the single chargino production cross section
$\sigma(e^+d_k\ra{\tilde\chi}^+_1 {\bar d}_j)$ for these two cases for $\lam'_{
1j1}=0.1$, and $M({\tilde u}_j)=210\gev$. For the pure Higgsino case, the cross
section is again highly suppressed. These decays can be neglected. For the
purely Wino case, $\sigma(e^+d_k\ra{\tilde\chi}^+_1 {\bar d}_j)\approx1\,pb$, 
for $M({\tilde\chi}^+_1)<180\gev$. This suppresses the NC-like production as in
the neutralino case and we obtain additional solutions. This is particularly
relevant for ${\tilde t }$ production.

The chargino decays to 
\beq
{\tilde \chi}^+_1\ra \left\{\nu_eu_j{\bar d}_k, \; {\bar\nu}_e{\bar u}_j{d}_k, 
\; e^-d_j{\bar d}_k,\;e^+{\bar d}_j{d}_k
\right\}.
\eeq
Again, the best search mode for a positron beam is the final state electron
which can be detected at HERA.

Summarising, the solutions (\ref{eq:solutions}.1,\ref{eq:solutions}.4,\ref{eq:solutions}.5)
as well as the special solution \eq{special1} can allow for substantial decays
to gauginos. The other solutions can only allow for Higgsino-like gauginos to
be lighter since the decay rates are negligible. Note that the new processes
via gaugino decays to positrons and neutrinos 
would be reconstructed at random $x$ and $Q^2$ for both NC and CC.

\section{Tests at HERA}
\labsec{hera}
There are several tests of the R-parity violating hypothesis which can be
performed at HERA. 
\begin{enumerate}
\item If HERA is switched back to $e^-p$ collisions the processes \eq{squark1}
-\eq{squark3} change to their charge conjugate
\barr
e^- + {u}_j &\ra& {\tilde { d}}_k\ra e^- + { u}_j,\lab{csquark1}\\
e^- +{ u}_j&\ra&{\tilde{d}}_k\ra{\nu_e}+{ d}_j, \lab{csquark2}\\
e^- + {\bar d}_k &\ra& {\tilde {\bar u}}_j \ra e^- +{\bar d}_k.
\lab{csquark3}
\earr
For $k=1$ or $j=1$ the event rates would increase or decrease, respectively by
\cite{mrs}
\barr
\begin{array}{ccc}
{{u}(x,Q^2)}/{{\bar u}(x,Q^2)}&\gsim& 100,\\ && \\
{{\bar d}(x,Q^2)}/{d(x,Q^2)}&\approx& 0.03,
\end{array} \quad {\rm for}\quad x=0.45, \;\;Q^2=4\cdot10^4\gev^2.
\earr
Using the $e^-P$
data from earlier runs at HERA, the first case would exclude any solution with
an incoming ${\bar u}$-quark. But no such solution was found due to stringent
low-energy bounds. For the solutions (\ref{eq:solutions}.1,\ref{eq:solutions}.4) this change
should lead to a non-observation. For the other solutions with higher
generation incoming quarks, there is no difference between the $q(x,Q^2)$
distribution and the ${\bar q}( x,Q^2)$ distribution and we expect no effect. 

\item In the operator $L_1Q_j\Dbar_k$, $L_1$ refers to the first generation 
lepton $SU(2)$ {\it doublet} superfield. Therefore, the positron in 
\eq{squark1}-\eq{squark3} is {\it right}-handed, and the electron in 
\eq{csquark1}-\eq{csquark3} is {\it left}-handed. Thus, when the 
electron/positron polarisers are installed the event rate should double or 
vanish depending on which polarisation is chosen for the lepton beam 
\cite{hewett}. This effect is independent of the quark generation indices.

\item Within supersymmetry with broken R-parity, the nature of the lightest
supersymmetric particle (LSP) is unknown. The decay spectrum of the produced 
squarks depends on the supersymmetry spectrum as a whole. In the previous 
Section, we considered the case where a neutralino or chargino is lighter
than the squark. This leads to an additional electron (positron) signal for 
a positron (electron) beam, which can be searched for.
\end{enumerate}
The most important conclusion from this is that HERA itself can determine the 
nature of the observed effect. The second conclusion we draw is that it is
essential for HERA to first continue running in the present mode and {\it not}
switch to an electron beam or run at lower energy. This way both experiments 
can establish whether there is a genuine effect or whether it is merely a 
statistical fluctuation. If HERA did switch to an electron beam and the effect
is due to the production of up-like squarks from down quarks the excess would
be decreased and require longer running time to establish its nature. We could
then end up in one year with two separate non-significant excesses and being
no-more the wiser.

\section{Signals at LEP~2}

In this Section we discuss three tests of the R-parity violation hypothesis at 
LEP~2. (1) Pair production of selectrons. The pair production of squarks at 
LEP~2 with $m_{\tilde q} = 200\gev$ is kinematically prohibited. Selectrons on 
the other hand, could be kinematically accessible at LEP without having been 
seen at HERA. In particular, they can decay through the same R-parity violating
operator $L_1Q_j\Dbar_k$. With the present data 
samples of $L\sim 25\pbinv$ per experiment at $\sqrt{s}=130-172
\gev$, selectron masses $m_{\tilde e} < 70\gev$ are accessible 
\cite{bartl.selectrons}. (2) Gauginos, if light, can be pair produced and
subsequently can decay via $L_1Q_j\Dbar_k$ to visible leptonic final states.
(3) Virtual t-channel exchange of squarks would give a contribution to the SM 
process $e^{+} e^{-} \ra q {\bar q}$ \cite{lqellis,choudhury.lep.virtual}. We
investigate these three effects in the following.

\subsection{Selectron pair production and the ALEPH 4-jet Anomaly}
If the selectrons are lighter than the gauginos, the dominant decay channel
for the Yukawa coupling size of interest  is:
\beq
{\tilde e}_{L,R} \ra u_j {\bar d}_k.
\eeq
Selectron pair production would then give rise to 4-jet signals at LEP~2 
through the process ${\tilde e_L}{\tilde e_R}\ra (u_j {\bar d}_k) ({\bar u}_j 
d_k)$. This scenario has been investigated in 
\cite{aleph.selectron.4jet.paper}, where it was found that selectron masses of 
$m_{{\tilde e}_L}=58\gev$, $m_{{\tilde e}_R}=48\gev$ could explain the
anomalous invariant mass peak of 4-jet final states observed by ALEPH 
\cite{ALEPH.reference}, as well as the apparent difference in mass $\Delta M 
\sim 10\gev$. It must be pointed out that the ALEPH anomaly is not seen by the
other three LEP experiments, and its origin is at present not understood. In 
this Subsection, we will assume that the ALEPH effect is due to new physics. 
It is then intriguing to ask whether the HERA excess when interpreted as 
s-channel squark production is compatible with the above interpretation of the 
ALEPH 4-jet signal. 
In order to explain the ALEPH data there are several requirements on
$\lam'_{1jk}$ \cite{aleph.selectron.4jet.paper}. 

(a) The solution to the ALEPH 4-jet anomaly requires the associated
production of an $SU(2)$ singlet selectron $({\tilde e}_R)$ and an $SU(2)$
doublet selectron $({\tilde e}_L)$ with different mass. The ${\tilde e}_R$ can 
not decay directly via the operator $L_1Q_j\Dbar_k$, only via the mixing with 
the ${\tilde e}_L$. If this mixing is small 
then for small $\lam'$ the ${\tilde e}_R$ will be long-lived. This in
turn can be observed in the ALEPH detector but wasn't. Thus the
coupling must be large: $\lam'_{1jk}>0.01$ \cite{aleph.selectron.4jet.paper}.

(b) The ALEPH 4-jet data excludes final states which contain b-quarks, and 
favours u,d,s quark final states; charm quarks are disfavoured, but not 
excluded. 
Including the tight constraints on $\lambda'_{111}$ one arrives at 
three solutions: $\lambda'_{112}, \,\lambda'_{121}\,\lam'_{122}$ ({\it c.f.}
Table~\ref{ALEPH.tab}). The solution with $\lambda'_{112}\gsim 0.01$ is favoured by the ALEPH data 
\cite{aleph.selectron.4jet.paper}. 

\begin{table}[h]
\centerline{\begin{tabular}{|l|l|}\hline&\\
Coupling &      Process which can explain\\  
& the ALEPH 4-jet anomaly\\
\hline
$\lambda'_{112} \gsim 0.01$& ${\tilde e}_L {\tilde e}_R  
\ra u {\bar u} s {\bar s}$ \\
$\lambda'_{121} > 0.01$& ${\tilde e}_L {\tilde e}_R  \ra c 
{\bar c} d {\bar d}$  \\
$\lambda'_{122} > 0.01$& ${\tilde e}_L {\tilde e}_R  \ra c 
{\bar c} s {\bar s}$ \\ \hline
\end{tabular}}
\caption[xx]{\label{ALEPH.tab} {\small Allowed couplings which can explain the ALEPH 4-jets.}}
\end{table}

As discussed in Section 4, the HERA high $Q^2$ excess, too, can be explained 
by one of the three couplings ({\it c.f.} Table~\ref{HERA.tab}). 

\begin{table}[h]
\centerline{\begin{tabular}{|l|l|l|}\hline &&\\
Coupling &      Process explaining the & Additional
signals  \\
& HERA high $Q^2$ anomaly & at HERA \\
\hline
$\lambda'_{121} > 0.04$& $e^+ d\ra {\tilde c} 
\ra e^+ d$ & $e^+ d \ra {\tilde c}
\ra {\tilde \chi}^0 u \hspace{0.5cm}(or \hspace{0.5cm} \tilde s {\tilde \chi}^+)$ \\
\hline 
\end{tabular}}
\caption[xx]{\label{HERA.tab} {\small Allowed coupling which can explain the HERA high
    $Q^2$ anomaly.}}
\end{table}

In order to fit the ALEPH data comfortably, the 
coupling $\lambda'_{121}$ should be as large as possible. This evades the 
effects of lifetime (which are now amplified by the presence of charm in the 
4-jet final states). But for  a large coupling (say 
$\lambda'_{121}=0.1$), the HERA excess cross-section of 0.17$\,pb$ only fits
the predicted cross-section if the squark coupling to the neutralino is large.
The additional decay mode ${\tilde c}\ra {\tilde\chi}^0 c$ must reduce the 
predicted cross-section $\sigma(e^+ {\bar u} \ra {\tilde c} \ra e^+ {\bar u})$
from $0.93\,pb$ to $0.17\,pb$ (Figure~\ref{sup.xsec.reduced}). A neutralino 
with  $M_{{\tilde \chi}^0}=125-165\gev$ which  couples with a dominant bino component
indeed fits the data. The additional process  $e^+ {\bar u} \ra {\tilde c} 
\ra {\tilde\chi}^0 c$ has a cross-section of $1.4\,pb$ (Figure 
\ref{neutralino.xsec}), which would predict around 50 events between the two 
experiment $H1$ and $ZEUS$ in the 1995/1996 data; a signal which should be 
observable. That cross-section decreases, as the value of 
$\lambda'_{121}$ is decreased, since the neutralino mass  has to be increased
correspondingly to fit the data\footnote{It was pointed out in 
\cite{altarelli} that cancellations in the gaugino coupling of the neutralino
to the left-handed charm squark can occur in large regions of Supersymmetric parameter
space. In this case decays of the squark to the chargino - which do not have the
same cancellations in coupling - would play the same role as the decays to the
neutralino discussed above.}.

We conclude that the ALEPH anomaly and the HERA excess can in principle be
simultaneously explained by the coupling $\lambda'_{121}>0.04$, although 
this scenario is less favoured by the ALEPH data.

\subsection{Other Signals at LEP~2}
Let us now  discuss the constraints placed on the supersymmetric spectrum
by the solutions of Eq.(\ref{eq:solutions}). We find four scenarios which determine
the gaugino spectrum:

\begin{enumerate}
\item[(a)] Solution (\ref{eq:solutions}.1) can accommodate gauginos which are 
substantially lighter
than $M_{\tilde q}$ and couple electroweakly to the squarks (e.g. non-Higgsino
like). 

\item[(b)] Solution (\ref{eq:solutions}.4) requires  $M_{{\tilde\chi}^0} > 
M_{\tilde t} - M_{t}$, and $M_{{\tilde \chi}^+} \gsim M_{\tilde t} -M_{b}$.

\item[(c)] All other cases require either large  gaugino masses ($
\gsim 200\gev$), which is of no interest to LEP~2, or

\item[(d)] they can accommodate light gauginos which a small coupling  
(e.g. Higgsino-like) to the squarks.
\end{enumerate}
In summary, we conclude that cases (a), (b) and (d) are  interesting for LEP~2
since charginos and neutralinos could be discovered up to the kinematic
limit of $\sqrt{s} / 2$. LEP~2 itself however has no way of testing the R-parity hypothesis of the
high $Q^2$ HERA excess, since the non-observation of R-parity violating SUSY 
at LEP~2 cannot rule out that the HERA effect is indeed a sign  of a s-channel
squark resonance. 

\subsection{Virtual Effects at LEP~2 from t-channel Squark Exchange}
\begin{figure}
\begin{center}
{\psfig{figure=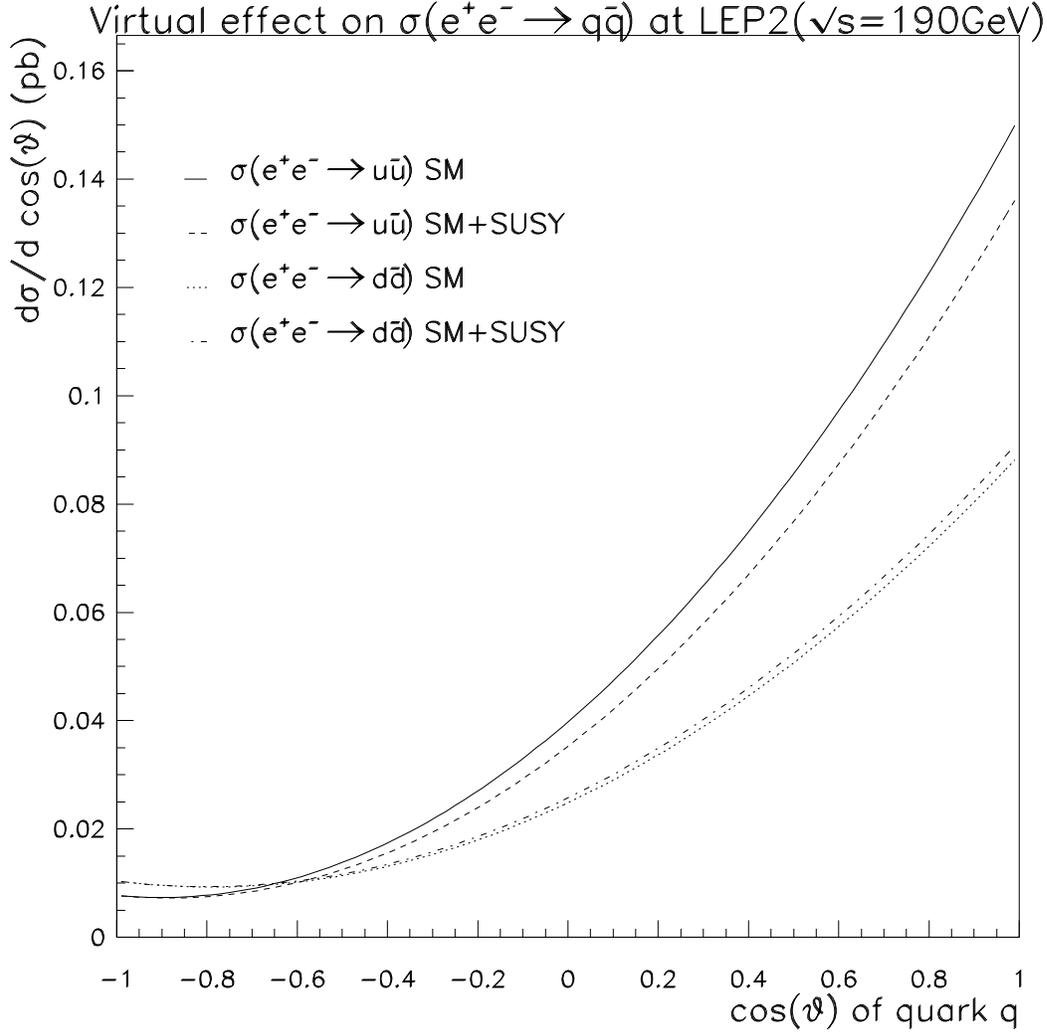,height=15cm}}
\end{center}
\label{lep.t.chan}
\caption[xx]{\small Differential cross-sections $d\sigma
/ d\cos{\theta}$ for $M_{{\tilde c}}$,$M_{\tilde s} = 200\gev$ and
$\lambda'=0.2$  respectively. We have included Initial
State Radiative corrections, and used the anti-ISR cut $\sqrt{{s'}/
{{s}}} > 0.9$ (see text).}
\label{lep.tchan.fig}
\end{figure}

We quote the cross-section  for this effect, which has also been studied in the literature \cite{lqellis,choudhury.lep.virtual},
 in the Appendix.  The magnitude of the effect is small, 
since the main contribution  to the cross-section 
is proportional to $\lam^{'2}$. Figure~\ref{lep.tchan.fig} 
shows the contribution of t-channel $\tilde c$ and $\tilde s$ exchange for 
$M_{\tilde q}=200\gev$ at $\sqrt{s}=190\gev$ including Initial State Radiation
corrections. Note, that
 the effect due to a  
$\tilde c$ exchange gives a positive contribution to $\sigma({e^+ e^- \ra 
d {\bar d}})$, while the $\tilde s$ exchange gives a negative contribution to 
$\sigma({e^+ e^- \ra u {\bar u}})$.

The overall effect on the total cross-section is shown in
Table~\ref{tab:lep.tchan.tab}. We have also included a column with 
cross-sections  including
an anti-ISR cut $\sqrt{{s'}/{s}}>0.9$, where s is the center of mass energy of
the incoming lepton beams, and $s'$ is the center of mass energy of the 
outgoing $q {\bar q}$ pair. 
The cut enhances the contribution of the t-channel squark exchange to the 
total $q {\bar q}$ cross-section. Nevertheless, the  effect
is still less than $3\%$ for a coupling $\lambda'=0.2$. 
LEP~2 can therefore only probe large Yukawa couplings.
For an integrated luminosity of  $400\pbinv$ (expected to be delivered to the four experiments for the 1997
run) LEP~2 would see a three sigma effect on the total $q {\bar q}$ cross-section for
couplings larger or equal than $\lambda' \gsim 0.2 (0.4)$ for $\tilde s (\tilde
c)$ squark exchange respectively.
 
Very recently, the OPAL collaboration have performed the measurement
\cite{opal}. For squark masses of $200\gev$, they are  sensitive
to couplings $\lam'\gsim0.4$, which is already on the verge of excluding 
the large coupling solutions in \eq{solutions}.

\begin{table}[t]
\begin{center}
\begin{tabular}{|c|c|c|c|}
\hline
Anti-ISR cut & $\sigma({e^+ e^- \ra d {\bar d}})$ 
        & $\sigma({e^+ e^- \ra d {\bar d}})$ & Effect on total cross-section \\
&       SM only& SM plus t-channel ${\tilde c}$ exchange & $\sigma({e^+ e^- \ra
q {\bar q}})$ \\
\hline
- &     18.62pb & 18.73pb & $+0.12\%$\\
$\sqrt{{s'}/{{s}}} > 0.9$ & 3.31pb &  3.42pb & $+0.53\%$\\ & & & \\
\hline
\hline
 & $\sigma({e^+ e^- \ra u {\bar u}})$ 
        & $\sigma({e^+ e^- \ra u {\bar u}})$ & \\
&       SM only& SM plus t-channel ${\tilde s}$ exchange & \\
\hline

- & 18.40pb & 17.77pb & $-0.68\%$\\
$\sqrt{{s'}/{{s}}} > 0.9$ & 5.30pb &  4.77pb & $-2.58\%$\\  
\hline
\end{tabular}
\end{center}
\protect\caption[xx]{\label{tab:lep.tchan.tab} 
{\small Cross-section values (including Initial State Radiation corrections)  for
the SM process and 
the SM plus t-channel squark exchange for $M_{\tilde q}=200\gev$ and
$\lambda=0.2$ at $\sqrt{s}=190\gev$. }}
\end{table}

\section{Signals at the Tevatron}
\labsec{fermilab}
There are several potential tests of the above `supersymmetry with broken 
R-parity' hypothesis (produce one squark via operator $L_1Q_j\Dbar_k$) at the 
Tevatron: (a) $t$-channel squark exchange interfering with Drell-Yan 
production, (b) single squark production, and (c) squark pair production.

(a) The operators $L_1Q_j\Dbar$ can also contribute to Drell-Yan production
via the $t$-channel exchange of a squark. The dominant effect is the 
interference with the SM. This has been studied in \cite{lqellis}. However, 
the above study did not consider the for us relevant range at the 
Tevatron, basically because it was not feasible with the projected luminosity.
In light of the planned Tevatron upgrade, we have repeated the analysis for
our solutions. However, even for the large couplings we do not find a 
measurable effect ($<5\%$) in this case.

(b) The scalar quarks can be singly produced via the parton-level processes
\barr
g+u_j&\ra& e^++ d_k, \\
g+d_k&\ra& e^-+ u_j,
\earr
as well as the complex conjugate production mechanisms. Here $g$ denotes a 
parton gluon. This is completely equivalent to single leptoquark production at 
the Tevatron which has been analysed in \cite{hewettlq}. The production cross 
section is proportional to $\lam^{'2}_{1jk}$. For $\lam'_{1jk}=0.1$ and $M_{LQ}
=200\gev$, we have $\sigma=0.02\,pb^{-1}$ \cite{hewettlq} for an incoming
first generation quark. Thus for our preferred solutions with $\lam'_{121},
\lam'_{131}=0.04\,(0.1)$ we do not expect any observable signal. For our more 
marginal solutions we have incoming sea-quarks which significantly suppresses
the cross section. Thus even for $\lam'=0.3,\,0.4$, we do not expect an
observable cross section.

(c) Squarks can be pair produced at the Tevatron via\footnote{The second 
process includes $t$-channel electron exchange via the R-parity violating 
operator. For the couplings we consider this contribution is negligible.} 
\barr
g+g&\ra& {\tilde q}+{\tilde {\bar q}}, \\
q+{\bar q} &\ra& {\tilde q}+{\tilde {\bar q}}.
\earr 
If the squark only decays via the dominant first lepton generation R-parity
violating operator $L_1Q_j\Dbar_k$ then this is equivalent to first generation
leptoquark production and decay. $D0$ have recently updated their analysis 
\cite{D0} and obtained preliminary lower bounds on such a 
leptoquark\footnote{These numbers are from March, 1997.}
\beq
D0:\quad
\begin{array}{cccccc}
M_{LQ}&>&175\gev,& {\rm for}\; BR(\phi_{LQ}\ra e^\pm q )&=&1, \\
M_{LQ}&>&147\gev,& {\rm for}\; BR(\phi_{LQ}\ra e^\pm q)&=&0.5,\\
M_{LQ}&>&71\gev,& {\rm for}\; BR(\phi_{LQ}\ra e^\pm q)&=&0.0.
\end{array}
\eeq
The first bound applies for a two-body leptonic branching ratio of 1, \ie the 
only decay mode is $\phi_{LQ}\ra e^\pm+q$. This can immediately be 
reinterpreted as a lower bound for up-like squarks, provided they can not 
decay via gauginos or other supersymmetric particles. The second, weaker bound 
applies to the case where there is a second decay without a charged lepton and 
which has equal decay width to the charged lepton decay. This can be 
reinterpreted as a lower bound for d-like squarks, since as shown in 
Eqs.\eqs{squark1}{squark2} they have two equal decay modes (provided the final 
state quark masses can be neglected).

Before applying these bounds we note that they only employ the tree-level
cross-section for leptoquark pair-production. In our case, for squark pair
production, the next-to-leading order calculation has been done. The production
cross section is increased by $\approx2$ \cite{kfac}. If we include this in the
$D0$-plot the more stringent bound increases by about $20\gev$ to $195\gev$. 
This should be further strengthened by combining the $D0$ and CDF leptoquark
bounds, and can most likely significantly cut into our required mass range
\eq{mass}. With the Tevatron upgrade the entire mass range can be covered.

For the down-like squarks, which have a decay branching ratio to charged
leptons $<0.5$ these $D0$ bounds are not relevant and our previous analysis of
Section 4.1 applies. For the up-like squarks with only R-parity violating decays,
solutions (\ref{eq:solutions}.1,\ref{eq:solutions}.2,\ref{eq:solutions}.4,\ref{eq:solutions}.5)
 and \eq{special1}, these bounds apply
directly and are relevant. At present they do not exclude any solution, but
after combining the numbers of {\it D0} and {\it CDF} the low-mass range of our
solutions could be excluded. In this case one must include the gaugino decays
of the squarks. As discussed in Section 4.2, the branching ratio
$BR({\tilde c}\ra e^++d)$ can be reduced to 0.1 for $\lam'_{121}=0.1$
and a neutralino mass $\approx165\gev$. In this case there are no
relevant bounds from the Tevatron.

\section{Conclusions}
\labsec{concl}
We have discussed the recently observed excess at HERA in the high $Q^2$ data
in the light of supersymmetry with broken R-parity. We have two solutions
which are in good agreement with all the data:
\begin{itemize}
\item $\lam'_{121}>0.04$ and the production of a scalar charm quark, ${\tilde 
c}$. If the ${\tilde c}$ is the LSP this solution can most likely be tested
at the Tevatron by combining all the existing data, otherwise with
the upgrade. If it is not the LSP,
then the ${\tilde c}$ can decay to a neutralino or a chargino. The resulting
spectrum can not be tested at the Tevatron with the leptoquark pair production
search. The neutralino production however should be testable at HERA
with a cross section of ${\cal O}(1\,pb)$. The signal is a charged
lepton of opposite sign to the beam lepton.

\item $\lam'_{131}>0.04$ and the production of a scalar top quark, stop,
${\tilde t}$. The possible solutions are analogous to the scalar charm with
the exception that the decay to the neutralino is kinematically prohibited
in most cases. In the renormalisation group studies in the MSSM a chargino
LSP is disfavoured. Again the stop-LSP solution can be tested at the Tevatron
and for a light chargino at HERA via chargino decay to an electron or positron.
\end{itemize}

We also have a set of further solutions which involve larger values of the
coupling. If some of the model dependent bounds we have discussed indeed hold,
these solutions are excluded. For the model independent indirect lower bounds
these solutions are still viable.

\begin{itemize}
\item $\lam'_{123}=0.4$ and the production of scalar charm quarks, ${\tilde c}$.
This solution prefers the lower mass range of the resonance solution to the
HERA data $(<210\gev)$. If the ${\tilde c}$ is the LSP, then the dilepton
search at the Tevatron can possibly test this model with the present data,
if both experiments are combined. It can most certainly test it with the 
upgrade. 

\item $\lam'_{123}=0.4$ and the production of scalar bottom quarks, ${\tilde b}
$. This solution also prefers the lower mass range $(<210\gev)$. It can not be
tested at the Tevatron since the charged leptonic branching ratio is less than
1/2. However this solution can be tested by the t-channel process at LEP~2.

\item $\lam'_{132}=0.3$ producing a ${\tilde t}$ squark. This solution allows 
for the full mass range. As for the ${\tilde c}$, this model can be tested
at the Tevatron if the squark is the LSP. If it is not the LSP there is room
in the coupling to allow for stop decays via the chargino (neutralino decays
are kinematically forbidden or suppressed by the top quark). 

\end{itemize}

We have one further solution: $\lam'_{123}=0.28$ and the production of two
kinds of squarks, a ${\tilde c}$ {\it and} a ${\tilde b}$. The tests are
analogous to those discussed above for the separate solutions $(\lam'_{123},\,
{\tilde c})$ and $(\lam'_{123},\, {\tilde b})$ .

It is amusing to note that in Ref.\cite{aliherbi} models with an anomaly-free 
family dependent $U(1)_R$ $R$-gauge symmetry were constructed which predicted 
the dominant R-parity violating operators $L_1Q_1\Dbar_2,$ and $L_1Q_2
\Dbar_1$.

We eagerly await further data to confirm or reject this hypothesis. It might
then also be possible to make a statement about CC DIS and contributions from 
supersymmetry decays.

{\bf Acknowledgements}

We thank Jon Butterworth and Ken Long for discussions of the ZEUS data. We
thank Sacha Davidson for discussions on indirect bounds. We thank Gian Giudice
and Michelangelo Mangano for pointing out an error in the cross section
calculation in the original version.

{\bf Note Added:} In our original manuscript we made a mistake in the cross
section computation. The correction has lead to changes in the allowed 
solutions. Since submitting our original manuscript several related 
papers have appeared \cite{altarelli,zerwas,others}. In \cite{altarelli,zerwas}
R-parity violation at HERA were also considered with similar conclusions to
ours. We have also taken the opportunity to update some of the indirect bounds
with more recent data and correct several typos in the formulas of the
Appendix.

\section{Appendix}

\subsection{Single Neutralino Production Cross-section at HERA}
The cross-sections for the process $e^-P \ra {\tilde \chi}^0_1 + X$ in the
approximation where the neutralino is a pure photino have been previously
calculated in \cite{jonherbi}. In this Section, we generalise the 
results to the case of a neutralino (which is an admixture of the photino, wino
and higgsino weak eigenstates).
The differential cross-section for 
single neutralino production via the R-parity violating
coupling $\lambda_{1jk}'$ at HERA may then be written as  
\barr
{\frac {d\sigma (e^+P \ra \chi^0 + X)} {dx dQ^2}} = 
{\frac {1} {16 \pi x^2 s^2}}
(&q_d(x,Q^2)|{\cal M}(e^+ d_k \ra \chi^0 u_j)|^2   +  \nonumber \\
 &{\bar q}_u(x,Q^2)|{\cal M}(e^+ {\bar u}_j \ra \chi^0 {\bar d}_k)|^2\,)
\label{eq:total.neut.xsec}
\earr
where $q_d(x,Q^2)$ and ${\bar q}_u(x,Q^2)$ give the probability of finding a 
$d_k$-quark and ${\bar u}_j$-quark respectively  inside the proton; $s,x,Q^2$
are the center of mass energy, the Bjorken scaling variable and the momentum
transfer squared. The Matrix elements can be obtained from 
\cite{peterherbi} by
crossing and are given by (neglecting initial state masses):
\barr
|{\cal M}(e^+ d_k \ra \chi^0 u_j)|^2&=&{\frac{g^2 \lambda'^2_{1jk}}{2}}\{  \nonumber \\
&&\frac {\shat} {(\shat-m^2_{\tilde uj})^2 + \Gamma^2_{\tilde uj}
m_{\tilde uj}^2} [(a(u_j)^2+b(u)^2)(\shat-m^2_{\chi^0}-m^2_{uj}) \nonumber\\
&&+4a(u_j)b(u)m_{uj}m_{\chi^0} ]  \nonumber\\
&+&\frac {m_{uj}^2-u} {(u-m^2_{\tilde dk})^2} [b({\bar
d})^2(m_{\chi^0}^2-u)]  \nonumber\\
&+&\frac {m_{uj}^2-u} {(t-m^2_{\tilde e})^2}
[b({e})^2(m_{\chi^0}^2-t)]  \nonumber\\
&-&\frac{\shat-m^2_{\tilde uj}}{[(\shat-m^2_{\tilde
uj})^2+\Gamma^2_{\tilde uj}m^2_{\tilde uj}][t-m^2_{\tilde
e}]}[2a(u_j)b(e)m_{uj}m_{\chi^0}\shat  \nonumber\\ 
&&+b(e)b(u)(\shat^2+t^2-u^2+(m_{\chi^0}^2+m_{uj}^2)(u-\shat-t))]  \nonumber\\
&+&\frac{1}{(t-m^2_{\tilde e})(u-m^2_{\tilde dk})}[b(e)b({\bar
d}) \nonumber \\
&&(t^2+u^2-\shat^2+(m^2_{\chi^0}+m^2_{uj})(\shat-t-u)+2m^2_{uj}m^2_{\chi^0})] \nonumber
\\
&+&\frac{\shat-m^2_{\tilde uj}}{[(\shat-m^2_{\tilde uj})^2+\Gamma^2_{\tilde
uj}m^2_{\tilde uj}][u-m^2_{\tilde dk}]}[2a(u_j)b(\bar
d)m_{uj}m_{\chi^0}\shat \nonumber\\
&&+b(u)b({\bar d})(\shat^2+u^2-t^2+(m^2_{\chi^0}+m^2_{uj})(t-\shat-u))]\}
\label{me.neut.d}
\earr
where  
$\shat,t,u$ are the Mandelstamm variables defined as $\shat=(p_e +
p_{dk})^2=xs$, $t=(p_e-p_{\chi^0})^2=-Q^2$, $u=m_{\chi^0}^2+m_{dk}^2-\shat-t$;
$g=\frac{e}{\sin{\theta_w}}$; $m_{\chi^0},m_{uj}$ are the masses of the final
state particles, and $m_{\tilde e},m_{\tilde uj},m_{\tilde dk}$ are the masses of
the exchanged scalar SUSY particles; $\Gamma_{\tilde uj}$ is the total width of
the scalar ${\tilde uj}$.

\barr
|{\cal M}(e^+ {\bar u}_j \ra \chi^0 {\bar d}_k)|^2&=&{\frac{g^2  \lambda'^2_{1jk}}{2}}\{
\nonumber \\
&& \frac{m_{dk}^2-u}{(u-m_{\tilde uj}^2)^2}b(u)^2(m_{\chi^0}^2-u)
\nonumber \\
&+&\frac{\shat}{(\shat-m_{\tilde dk}^2)^2 + \Gamma^2_{\tilde dk} m^2_{\tilde
dk}}[(a(d_k)^2+b({\bar d})^2)(\shat-m_{\chi^0}^2-m^2_{dk}) \nonumber \\
&&-4a(d_k)b({\bar d})m_{dk}m_{\chi^0}] \nonumber \\
&+&\frac{m^2_d - t}{(t-m^2_{\tilde e})^2}b(e)^2(m^2_{\chi^0}-t)
\nonumber\\
&-&\frac{1}{(t-m^2_{\tilde e})(u-m^2_{\tilde
u})}b(e)b(u)[t^2-s^2+u^2+(m^2_{\chi^0}+m^2_{dk})(s-t-u)+2m^2_{dk}m^2_{\chi^0}]
\nonumber \\
&-&\frac{\shat-m^2_{\tilde dk}}{[(\shat-m^2_{\tilde dk})^2+\Gamma^2_{\tilde
dk}m^2_{\tilde dk}][u-m^2_{\tilde uj}]} [2a(d_k)b(u)m_{dk}m_{\chi^0}\shat
\nonumber \\
&&-b(u)b({\bar d})(u^2+s^2-t^2+(m^2_{\chi^0}+m_{dk}^2)(t-s-u))] \nonumber \\
&-&\frac{\shat-m^2_{\tilde dk}}{[(\shat-m^2_{\tilde dk})^2+\Gamma^2_{\tilde
dk}m^2_{\tilde dk}][t-m_{\tilde d}^2]}[2a(d_k)b(e)m_{dk}m_{\chi^0}\shat
\nonumber \\
&&- b(e)b({\bar d})(s^2+t^2-u^2+(m_{\chi^0}^2=m^2_{dk})(u-t-s))]\}
\label{me.neut.ubar}
\earr
where $\shat,t,u$ are now defined as $\shat=(p_e +
p_{uj})^2=xs$, $t=(p_e-p_{\chi^0})^2=-Q^2$,
$u=m_{\chi^0}^2+m_{uj}^2-\shat-t$. The total cross-section can be obtained from
Eq.(\ref{eq:total.neut.xsec}) by
integrating over the $x,Q^2$ range
\barr
x_{min}&=&(m_{\chi^0}^2+m_{fs}^2)/s \nonumber \\
x_{max}&=&1 \nonumber \\
Q^2_{min}&=&\shat-m_{\chi^0}^2 \nonumber \\
Q^2_{max}&=&\shat 
\earr
and $m_{fs}$ is the mass of the final state quark, i.e. $m_{uj}$, $m_{dk}$ for
Eqns. (\ref{me.neut.d}),(\ref{me.neut.ubar}) respectively. The coupling 
constants a,b are given in
{Table~A.1 of reference \cite{peterherbi}. The cross-section for the
process $e^- P \ra \chi^0 + X$ can be obtained from the above result by charge
conjugation:
\barr
{\frac {d\sigma (e^-P \ra \chi^0 + X)} {dx dQ^2}} = 
{\frac {1} {16 \pi x^2 s^2}}
(&{\bar q}_d(x,Q^2)|{\cal M}(e^- {\bar d}_k \ra \chi^0 {\bar u}_j)|^2   +  \nonumber \\
 &{q}_u(x,Q^2)|{\cal M}(e^- {u}_j \ra \chi^0 {d}_k)|^2)
\earr
where $|{\cal M}(e^- {\bar d}_k \ra \chi^0 {\bar u}_j)|^2=|{\cal M}(e^+ d_k \ra \chi^0 u_j)|^2$
and $|{\cal M}(e^- {u}_j \ra \chi^0 {d}_k)|^2=|{\cal M}(e^+ {\bar u}_j \ra \chi^0 {\bar d}_k)|^2$.

\subsection{Single Chargino Production Cross-section at HERA}
The differential cross-section for single chargino production via the R-parity violating
coupling $\lambda_{1jk}'$ at HERA may be written as  
\barr
{\frac {d\sigma (e^+P \ra \chi^+ + X)} {dx dQ^2}} = 
{\frac {1} {16 \pi x^2 s^2}}
(&q_d(x,Q^2)|{\cal M}(e^+ d_k \ra \chi^+ d_j)|^2   +   \nonumber \\
 &{\bar q}_u(x,Q^2)|{\cal M}(e^+ {\bar u}_j \ra \chi^+ {\bar u}_k)|^2)
\earr
where the Matrix elements can be obtained from \cite{peterherbi2} by
crossing and are given by 
\barr
|{\cal M}(e^+ d_k \ra \chi^+ d_j)|^2&=&{\frac {g^2 \lambda'^2_{1jk}} {4}}\{
\nonumber \\
&&{\frac{m^2_{dj}-t}{(t-m^2_{\tilde
\nu})^2}}(\gamma_L^2+\gamma_R^2)(m^2_{\chi^+}-t) \nonumber\\
&+&{\frac{\shat}{(\shat^2-m_{\tilde uj}^2)^2+\Gamma^2_{\tilde uj}m^2_{\tilde
uj}}}[(\delta_L^2+\delta_R^2)(\shat-m^2_{\chi^+}-m^2_{dj})+8{\cal R}e\{\delta_L
\delta^*_R m_{dj} m_{\chi^+}\}] \nonumber\\
&-&{\frac{\shat-m^2_{\tilde uj}}{[(\shat-m^2_{\tilde uj})^2 + \Gamma^2_{\tilde
uj}m^2_{\tilde uj}][t-m^2_{\tilde \nu}]}} 
{\cal R}e \{\gamma_L \delta_L^*
[\shat^2+t^2-u^2 \nonumber \\
&& +(m_{dj}^2 +m^2_{\chi^+})(u-s-t)] + 2 \gamma_L \delta_R^* m_{dj}
M_{\chi^+} \shat\}\}
\earr
where $\shat,t,u$ are defined by $\shat=(p_e +
p_{dk})^2=xs$, $t=(p_e-p_{\chi^+})^2=-Q^2$,
$u=m_{\chi^+}^2+m_{dj}^2-\shat-t$. And 
\barr
|{\cal M}(e^+ {\bar u}_j \ra \chi^+ {\bar u}_k)|^2 = 
\frac{g^2 \lambda'^2_{1jk}} {4} \frac{1}{(\shat^2-m^2_{\tilde dk})^2 + \Gamma^2_{\tilde dk} m^2_{\tilde
dk}} \epsilon_R^2 (\shat^2 - \shat m_{\chi^+}^2 - \shat m^2_{uk})
\earr
where $\shat,t,u$ are defined by $\shat=(p_e +
p_{uj})^2=xs$, $t=(p_e-p_{\chi^+})^2=-Q^2$,
$u=m_{\chi^+}^2+m_{uk}^2-\shat-t$ and 
the coupling constants $\gamma,\delta,\epsilon$  are 
\barr
\gamma_L =  i V^*_{12} & , & \epsilon_R  =  - \frac {im_{dk} U_{12}}{\sqrt{2} M_W
\cos{\beta}} \nonumber \\
\delta_L  =  \gamma_L & , & \delta_R  =  - \frac{im_{dj} U_{12}}{\sqrt{2} M_W
\cos{\beta}}.
\earr
We follow here the notation of \cite{gunhab}, where one can  find the 
expressions for the matrices $U_{ij},\,V_{ij}$ which diagonalise the
chargino mass matrix.
 
\subsection{Virtual Squark t-channel Exchange at LEP}
The differential cross-section can  be expressed as  
\barr
\frac{d\sigma(e^+ e^- \ra q \bar{q})}{d\cos{\theta}} = \frac
{d\sigma_{SM}}{d\cos{\theta}} + \frac{3}{32 \pi s}(A_{1} + A_{2})
\earr 
and the amplitudes are given by 
\barr
A_1&=& \frac{\lambda'^4}{4(t-\tilde{m}^2)^2} t^2 \nonumber\\
A_2&=&-\frac{\lambda'^2 t^2}{(t-\tilde{m}^2)} (\frac{e^2 Q_e Q_q}{s} + 
{\cal R}e\{\frac{g_Z^2 a_e a_q }{s-M_Z^2 + i\Gamma_Z M_Z}\}) 
\earr
$A_1,A_2$ correspond to the contributions from the
t-channel diagram and the SM interference respectively. Here 
${\tilde m}$ is the mass of the exchanged squark; $Q_e,Q_q$ are the electric
charge of the electron and quark $q$ respectively;
$g_Z=\frac{e}{sin{\theta_w}cos{\theta_w}}$;  $\theta$ is the angle between the
incoming electron and the quark $q$;  $t=-\frac{s}{2}(1+\cos{\theta})$; and
$a_e,a_q$ are coupling constants defined by
\barr
a_e &=& - {\small \frac{1}{2}}  + \sin^2\theta_w \nonumber\\
a_u &=& {\small \frac{1}{2}}- {\small \frac{2}{3}} \sin^2\theta_w \nonumber\\
a_d &=& {\small \frac{1}{3}}\sin^2\theta_w
\earr


\begin{thebibliography}{99}
\bibitem{us}{We have extracted this number from the tables in \cite{H1,ZEUS}.}
\end{thebibliography}
\end{document}